\documentclass[12pt]{article}

\usepackage{epsf,amsfonts,hyperref}
\renewcommand{\appendix}[1]{
    \addtocounter{section}{1}
    \setcounter{equation}{0}
    \renewcommand{\thesection}{\Alph{section}}
    \section*{Appendix \thesection\protect\indent #1}
    \addcontentsline{toc}{section}{Appendix \thesection\ \ \ #1}
}
\newcommand\encadremath[1]{\vbox{\hrule\hbox{\vrule\kern8pt
\vbox{\kern8pt \hbox{$\displaystyle #1$}\kern8pt}
\kern8pt\vrule}\hrule}}
\def\enca#1{\vbox{\hrule\hbox{
\vrule\kern8pt\vbox{\kern8pt \hbox{$\displaystyle #1$}
\kern8pt} \kern8pt\vrule}\hrule}}

\newcommand\figureframex[3]{
\begin{figure}[bth]
\hrule\hbox{\vrule\kern8pt
\vbox{\kern8pt \vbox{
\begin{center}
{\mbox{\epsfxsize=#1.truecm\epsfbox{#2}}}
\end{center}
\caption{#3}
}\kern8pt}
\kern8pt\vrule}\hrule
\end{figure}
}
\newcommand\figureframey[3]{
\begin{figure}[bth]
\hrule\hbox{\vrule\kern8pt
\vbox{\kern8pt \vbox{
\begin{center}
{\mbox{\epsfysize=#1.truecm\epsfbox{#2}}}
\end{center}
\caption{#3}
}\kern8pt}
\kern8pt\vrule}\hrule\end{figure}
}

\renewcommand{\thesection}{\arabic{section}}

\makeatletter
\@addtoreset{equation}{section}
\makeatother
\newtheorem{theorem}{Theorem}[section]

\newtheorem{remark}{Remark}[section]
\newtheorem{proposition}{Proposition}[section]
\newtheorem{lemma}{Lemma}[section]
\newtheorem{corollary}{Corollary}[section]
\newtheorem{definition}{Definition}[section]
\newcommand{\eq}[1]{eq.~(\ref{#1})}
\def\br{\begin{remark}\rm\small}
\def\er{\end{remark}}
\def\bt{\begin{theorem}}
\def\et{\end{theorem}}
\def\bd{\begin{definition}}
\def\ed{\end{definition}}
\def\bp{\begin{proposition}}
\def\ep{\end{proposition}}
\def\bl{\begin{lemma}}
\def\el{\end{lemma}}
\def\bc{\begin{corollary}}
\def\ec{\end{corollary}}
\def\beaq{\begin{eqnarray}}
\def\eeaq{\end{eqnarray}}
\newcommand{\proof}[1]{{\noindent \bf proof:}\par
{#1} $\square$}

\newcommand{\refeq}[1]{eq.(\ref{#1})}

\newcommand{\beq}{\begin{equation}}
\newcommand{\eeq}{\end{equation}}
\newcommand{\bea}{\begin{eqnarray}}
\newcommand{\eea}{\end{eqnarray}}

\renewcommand{\and}{{\qquad {\rm and} \qquad}}

\newcommand{\virg}{{\qquad , \qquad}}

 \newcommand{\Tr}{{\,\rm Tr}\:}

\newcommand{\Res}{\mathop{\,\rm Res\,}}

\newcommand{\td}[1]{{\tilde{#1}}}

\newcommand{\om}{\omega}
\newcommand{\Om}{\Omega}

\newcommand{\ii}{{\mathrm{i}}}

\newcommand{\ee}[1]{{{\rm e}^{#1}}}

\renewcommand{\d}{{{\partial}}}

\newcommand{\Pint}{{\int\kern -1.em -\kern-.25em}}

\renewcommand{\Re}{{\mathrm{Re}}}
\renewcommand{\Im}{{\mathrm{Im}}}

\renewcommand{\L}{\Lambda}

\newcommand{\curve}{{\cal C}}
\newcommand{\bcycle}{{\cal B}}
\newcommand{\acycle}{{\cal A}}

\newcommand{\genus}{{\mathfrak g}}

\title{{\bf Algebraic variational principle for the spectral curve of matrix models} \vspace{.5cm}}
\author{
{\bf B. Eynard},
{{\small {\it Institut de Physique Th\'{e}orique, CEA Saclay,}
{\it CRM, Centre de Recherche Math\'ematiques, Montr\'eal QC Canada}}}
}

\begin{document}

\pagestyle{empty}
\hfill IPHT T-14/037\\
\indent \hfill CRM-2014
\addtolength{\baselineskip}{0.20\baselineskip}
\baselineskip 16pt 
\begin{center}
\begin{Large}\fontfamily{cmss}
\fontsize{20pt}{30pt}
\selectfont
\medskip
\textbf{Another algebraic variational principle for the spectral curve of matrix models}
\end{Large}\\
\bigskip
\bigskip
{\sl B.\ Eynard}\hspace*{0.05cm}${}^{1\,2}$\\
${}^1$ Institut de physique th\'{e}orique, 
F-91191 Gif-sur-Yvette Cedex, France.\\
${}^2$ CRM, Centre de Recherche Math\'ematiques, Montr\'eal QC Canada.\\
\vspace{6pt}
\end{center}

\vspace{20pt}
\begin{center}
{\bf Abstract}

We propose an alternative variational principle whose critical point is the algebraic plane curve associated to a matrix model (the spectral curve, i.e. the large $N$ limit of the resolvent). More generally, we consider a variational principle that is equivalent to the problem of finding a plane curve with given asymptotics and given cycle integrals.
This variational principle is not given by extremization of the energy, but by the extremization of an "entropy".

\end{center}


%
%
%
%
%
%
%
%

%







\section{Introduction}

To a random matrix model is associated an algebraic curve, often called "spectral curve". Most often this is the Stieljes transform of the "equilibrium spectral density", although not always.
That algebraic curve is either obtained from the large N limit of the loop equations, or the large N limit of the saddle point equation, see for instance the review \cite{ZJDFG}.
It is a curve with some specific type of singularities and boundary conditions.

\medskip
It has been known for long, in many cases, that the large $N$ density of eigenvalues can be found by extremizing an energy functional in the space of measures, and it turns out that the extremal measure is an algebraic function.

\medskip

Our goal in this article, is to present another (in fact several other) variational principle, yielding the same spectral curve, but by extremizing only a functional in the space of algebraic curves (not using measures).

\section{1- matrix model}

\subsection{Introduction to random matrices}

Consider a random hermitian matrix $M$ of size $N$ (see \cite{Mehta}), with probability law:
\beq
\frac{1}{Z}\,\ee{-\frac{N}{t}\,\Tr V(M)}\,\,dM
\eeq
where $
dM = \prod_i dM_{i,i}\,\prod_{i<j}\,d\Re\,M_{i,j}\,\,d\Im\,M_{i,j}
$ is the $U(N)$ invariant Lebesgue measure on $H_N$,
and where $V(x) = \sum_k \frac{t_k}{k} x^k$ is a polynomial called the "potential", and $t>0$ is often called "temperature".
The normalization factor $Z$ is called the partition function:
\beq\label{defZ1MM}
Z = \int_{H_N}\, \ee{-\frac{N}{t}\,\Tr V(M)}\,\,dM.
\eeq
One can also extend this, and replace random hermitian matrices, by random "normal matrices with eigenvalues on some contour $\Gamma$":
\beq
H_N(\Gamma) = \{M=U\Lambda U^\dagger\,\,\,|\,U\in U(N)\,,\,\Lambda={\rm diag}(\lambda_1,\dots,\lambda_N)\, , \, \lambda_i\in \Gamma\}
\eeq
equipped with the measure $dM = \prod_{i<j} (\lambda_i-\lambda_j)^2\,dU\,\,\prod_i d\lambda_i$ where $dU$ is the Haar measuer on $U(N)$ and $d\lambda_i$ is the curvilinear measure along $\Gamma$.
For instance when $\Gamma=\mathbb R$ this coincides with Hermitian matrices:
\beq
H_N(\mathbb R)=H_N\,\, , \,\, dM={\rm Lebesgue\,measure\,on}\,H_N,
\eeq
and when $\Gamma=S^1=$unit circle in $\mathbb C$, this coincides with the "circular ensemble" $U(N)$ with its Haar measure:
\beq
H_N(S^1)=U(N)\,\, , \,\, (\det M)^{-N} dM={\rm Haar\,measure\,on}\,U(N).
\eeq

The expectation value of the resolvent:
\beq\label{defW}
W(x)  = \frac{t}{N}\,\mathbb E \left( \Tr \,(x-M)^{-1} \right)
\eeq
plays an important role, indeed its singularities  encode the information on the spectrum of $M$.

In many cases (depending on the choice of potential $V$ and on the choice of contour $\Gamma$), it is known (see \cite{Johanson, BGK13, Mehta} for instance), that $W(x)$ has a large $N$ limit:
\beq
W(x) \mathop{{\sim}}_{N\to\infty} \om(x)
\eeq
and in many cases (again depending on the choice of potential $V$ and contour), it is an algebraic function of $x$, i.e. it satisfies an algebraic equation:
\beq
P(x,\om(x))=0
\qquad ,\,\,
P(x,y) = \sum_{i,j} P_{i,j} x^i\,y^j.
\eeq
This algebraic equation has several solutions (several branches) $y=Y_k(x)$, $k=1,\dots,d$ where $d=\deg_y P$, and $\om(x)=Y_0(x)$ is only one branch (it has to be a branch which behaves as $\om(x)\sim t/x$ at large $x$, due to \refeq{defW}).
Alternatively, one can view $\om(x)$ as a multivalued function, or alternatively, it can be viewed as a meromorphic function on the compact Riemann surface $\curve$ defined by the algebraic equation $P(x,y)=0$.

For the 1-matrix model, the polynomial $P(x,y)$ is always quadratic in $y$ (the algebraic equation is said to be "hyperelliptical"), and always of the form:
\beq
P(x,y) = y^2 - y V'(x) + P(x)
\eeq
Finding $\om(x)=y$ amounts to finding the polynomial $P(x)$.

Since there is a branch of $\om(x)$ which begaves as $t/x$ at large $x$, this implies that $P(x)\sim tV'(x)/x$ at large $x$, i.e. $P(x)$ has degree $\deg V'-1$.

Then we have:
\beq
\om(x) = y = \frac{1}{2}\left( V'(x) \pm \sqrt{V'(x)^2-4 P(x)}\right).
\eeq
Branchcuts occur at the odd zeroes of $U(x)=V'(x)^2-4 P(x)$. Since $U(x)$ has even degree, there is necessarily an even number of odd zeroes, say $2s+2$ odd zeroes.

Let us denote:
\beq
U(x) = V'(x)^2-4 P(x) = M(x)^2\,\sigma(x)
\eeq
\beq
\sigma(x) = \prod_{k=1}^{2s+2} (x-a_k) = {\rm  product\,of\,odd\,zeroes}\, , \quad M(x) = \sqrt{\frac{U(x)}{\sigma(x)}} =  {\rm  product\,of\,even\,zeroes}.
\eeq
The points $a_k$ are called the branchpoints.

\subsubsection{Filling fractions}

Let us define for $\alpha=1,\dots,s$:
\beq
\acycle_\alpha={\rm clockwise\,contour \, surrounding}\,\, [a_{2_\alpha-1},a_{2\alpha}].
\eeq

Very often, it is interesting to consider matrix models with "fixed filling fractions", i.e. where the number of eigenvalues of $M$ in a certain region of the complex plane is held fixed.
The number $n_\alpha$ of eigenvalues of $M$ enclosed by a clockwise contour $C_\alpha$ is:
\beq
n_\alpha = -\,\frac{N}{2i\pi\,t}\,\oint_{C_\alpha}\,W(x)\,dx
\eeq
In the large $N$ limit, the fixed filling fraction condition amounts to fix:
\beq
\frac{t\,n_\alpha}{N} =-\,\frac{1}{2i\pi}\,\oint_{\acycle_\alpha}\,\om(x)\,dx = \epsilon_\alpha .
\eeq
The numbers $\epsilon_\alpha$ are called "filling fractions", they tell the number (times $t/N$)  of eigenvalues of $M$ which concentrate along the segment $[a_{2_\alpha-1},a_{2\alpha}]$.

\subsubsection{Loop equations}

Our goal now is to find the polynomial $P(x)$, as a function of the potential $V(x)$, the contour $\Gamma$ and the filling fractions $\epsilon_\alpha$'s.

It is well known that this polynomial can be determined by the following equations \cite{ZJDFG, Virasoro}:

\bd[Loop equations]

The loop equations of the 1-matrix model with potential $V$ and with filling fractions $\epsilon_\alpha$ is the following set of equations:
\beq\label{loopeq1M1}
\encadremath{
\left\{\begin{array}{l}
\exists\,{\rm polynomial}\, P(x)\,\,\, {\rm such\,that}\,\, \om^2(x)-\om(x)\,V'(x)+P(x) = 0 \cr
{} \cr
\exists\, {\rm branch}\quad \om(x)\mathop{{\sim}}_{x\to\infty} t/x +O(1/x^2) \cr
{} \cr
\forall \,\alpha=1,\dots,s \, , \qquad  -\,\frac{1}{2i\pi}\,\oint_{\acycle_\alpha}\,\om(x)\,dx = \epsilon_\alpha
\end{array}\right.
}\eeq

\ed

Let us check that indeed this system implies as many equations as unknowns:
let $d=\deg V'$. 
Observe that the second equation ($\om\sim t/x$) implies that $\deg P= d-1$, and this equation also fixes the leading coefficient of $P(x)$, it gives:
\beq
\lim_{x\to\infty} \frac{xP(x) }{V'(x)} = t.
\eeq
$P(x)$ has thus $d-1$ unknown coefficients.
The constraint that $U(x) = V'(x)^2-4P(x)$ has only $2s+2$ odd zeroes, i.e. $d-s-1$ even zeroes, imposes $d-s-1$ additional constraints on $P(x)$, i.e. there are only $s$ unknown coefficients left in $P(x)$.
Those $s$ coefficients are then determined by the $s$ filling fraction equations.

\medskip
Our goal is not to study those equations, in particular the existence and unicity or not of solutions, as there already is a large literature about them, but to show that the same equation \eq{loopeq1M1} can be obtained from a local variational principle.

\subsubsection{Usual energy variational principle}

In the case where $M$ is a hermitian matrix (eigenvalues $\in\mathbb R$), and $V(x)$ is a real potential bounded from below on $\mathbb R$, 
there is a known variational principle to find $\om(x)$. 
$\om(x)$ is the Stieljes transform of a positive measure $d\rho(x)$ on $\mathbb R$, such that:
\beq
\om(x) = \int_{x'\in{\rm supp.}\,d\rho} \frac{d\rho(x')}{x-x'}
\qquad , \quad
2\ii \pi\,\,\frac{d\rho(x)}{dx} = \om(x-i0)-\om(x+i0).
\eeq
It is well known that the measure $d\rho$ can be found as the {\bf unique}  minimum of the convex functional on the space of measures $d\rho$:
\bea
{\cal S}[d\rho]
&=& \int_{x\in{\rm supp.}\,d\rho} V(x)\,d\rho(x) - \int_{x\in{\rm supp.}\,d\rho} \int_{x'\in{\rm supp.}\,d\rho} \,d\rho(x)\,d\rho(x')\,\ln{|x-x'|} \cr
&& + \sum_\alpha \eta_\alpha\,\int_{a_{2\alpha-1}}^{a_{2\alpha}} d\rho(x)
\eea
where $\eta_\alpha$ are Lagrange multipliers determined by requiring that
\beq
\int_{a_{2\alpha-1}}^{a_{2\alpha}} d\rho(x) = \epsilon_\alpha.
\eeq

This functional is convex when $V$ is real, ${\rm supp.}d\rho\subset \mathbb R$  and $d\rho>0$, so that this variational problem can  be proved to have a unique minimum, and one finds that the minimum $d\rho$ is algebraic $d\rho(x)=\frac{1}{\pi}\sqrt{4P(x)-V'^2(x)}\,dx$, and is solution of the loop equations above.

\medskip
In case $V$ is not real, or $\Gamma\neq \mathbb R$ or $d\rho$ is not a positive measure on $\mathbb R$, usually the support of $d\rho$ is also  unknown (free frontier problem), and the above functional is then no longer convex, instead of an extremum, it has a saddle--point, and it is not known {\em in general} whether saddle--points are unique or not (it might be known case by case).
However, in all cases, any continuous saddle--point of the functional ${\cal S}$ is a solution to loop equations, and vice/versa, any solution of loop equations is a saddle--point of ${\cal S}$.

\medskip
Our purpose here is to propose another variational principle.

\subsection{New variational principle}

\subsubsection{Algebro geometric notations}

Consider a 2-sheeted hyperelliptical Riemann surface.
Its complex structure is determined by the location of its branch points $a_\alpha$, $\alpha=1,\dots,2s+2$, as well as a choice of non--intersecting paths joining them, of the form:
\beq
\acycle_\alpha={\rm counter-clockwise\,contour\,around}\,\,[a_{2_\alpha-1},a_{2\alpha}]
\virg
\bcycle_\alpha=[a_{2_\alpha},a_{2s+1}]
\eeq
so that
\beq
\acycle_\alpha\cap\bcycle_\beta=\delta_{\alpha,\beta}.
\eeq

Define:
\beq
\sigma(x)=\prod_{\alpha=1}^{2s+2} (x-a_\alpha)
\eeq

Define the "Cauchy kernel":
\beq
dS(x) =  {x^{s}+P_{s-1}(x)\over \sqrt{\sigma(x)}}\,dx
\eeq
where $P_{s-1}$ is the unique polynomial of degree $s-1$, whose $s$ coefficients are uniquely determined by:
\beq
\forall \alpha=1,\dots,s,\qquad \quad \int_{\acycle_\alpha} dS =0
\eeq
Indeed, this system of equation is linear in the coefficients of $P_{s-1}$ and admits a unique solution\footnote{The fact that this linear system has a unique solution is a standard result in the theory of Riemann surfaces, see \cite{Farkas, Fay}. It can be seen as a consequence of Riemann-Roch theorem.}.

We define:
\beq
\L(x)=\int_{a_{2s+2}}^x dS
\eeq
and since $dS\sim \pm \frac{dx}{x}$ at $x\to\infty_\pm$ we may define:
\beq
\gamma=\mathop{{\rm lim\,}}_{\infty_+} x/\L(x).
\eeq

We also define the "fundamental 2nd kind form":
\beq
B(x,x')={dx dx'   \,(\sqrt{\sigma(x)}+\sqrt{\sigma(x')})^2\over 4(x-x')^2\, \sqrt{\sigma(x)}\,\sqrt{\sigma(x')}} + {dx dx' P(x,x')\over  \sqrt{\sigma(x)}\,\sqrt{\sigma(x')}} = B(x',x)
\eeq
where $P(x,x')$ is the unique\footnote{Again, existence and unicity of such $B$ is a classical result of Riemannian geometry \cite{Fay}.} symmetric polynomial in $x$ and $x'$ of degree $s-1$, determined by:
\beq
\forall \alpha=1,\dots,s,\,\, \forall x,\qquad \quad \int_{x'\in \acycle_\alpha} B(x,x') =0
\eeq
We have:
\beq
dS(x)= -\,\frac{dx}{2x} +  \int_{x'=\infty_-}^{\infty_+} B(x,x')
\eeq

The holomorphic forms $du_i(x)$ are defined as:
\beq
du_i(x) = \frac{1}{2\i \pi}\,\oint_{x'\in\bcycle_i} B(x,x') = \frac{L_i(x)\,dx}{\sqrt{\sigma(x)}}
\eeq
where $L_i(x)$ is the unique polynomial of degree $\leq s-1$ such that
\beq
\oint_{x\in\acycle_i} du_j(x) =\delta_{i,j}.
\eeq

\subsubsection{The Variational principle}

Consider the following functional:
\begin{definition}
Let $t>0$ and $V'(x)=\sum_{k=1}^d t_k x^{k-1}$ be a given potential, and let $s\leq d$ be an integer, and $\epsilon_{i}$, $i=1,\dots,s$ be given filling fractions. For any hyperelliptical surface of genus $s$ with branch points $a_1,\dots, a_{2s+2}$, we define: 
\bea
\mu(\{t_k\},t;\{\epsilon\},\{a_\alpha\})
&:=& - \sum_k \frac{t_k}{k} \Res_{\infty_-} x^k dS
+ \sum_{i=1}^s \epsilon_i \oint_{\bcycle_i} dS
- 2t\ln{\gamma} \cr
\eea
\end{definition}

It is such that the variational principle $d\mu=0$ is equivalent to loop equations \eq{loopeq1M1}, i.e. the following theorem:
\bt
The set of equations
\beq
\forall\,\alpha=1,\dots,2s+2\,\,\, , \qquad 
\frac{\d \mu}{\d a_\alpha}=0
\eeq
is equivalent to the loop equations \eq{loopeq1M1}.

\et

\proof{
We have the Rauch variational formula \cite{Farkas, Fay}:
\beq
{\d B(p,q)\over d a_\alpha} = \Res_{\zeta\to a_\alpha} {B(p,\zeta)B(q,\zeta)\over dx(\zeta)}
\eeq
thus:
\beq
{\d dS(p)\over d a_\alpha} = \Res_{\zeta\to a_\alpha} {B(p,\zeta)dS(\zeta)\over dx(\zeta)}
\eeq
\beq
{\d \ln\gamma^2\over d a_\alpha} = -\Res_{\zeta\to a_\alpha} {dS(\zeta)dS(\zeta)\over dx(\zeta)}
\eeq
where $\zeta$ is a local coordinate on the Riemann surface, and residues are taken on the Riemann surface. For instance near a branchpoint $a_\alpha$, a good coordinate is $\zeta=\sqrt{x-a_\alpha}$. By abuse of notation we identify the point $\zeta(a_\alpha)\equiv a_\alpha$ with its $x$ value $a_\alpha=x(\zeta(a_\alpha))$.

The differential form $dx$ has a zeroe at $a_\alpha$, as can be seen from the choice of local coordinate $x=a_\alpha+\zeta^2$, for which $dx = 2\zeta\,d\zeta$, which vanishes at $\zeta=0$.

Thus:
\beq
{\d \mu\over d a_\alpha}
= - \Res_{\zeta\to \zeta(a_\alpha)} {dS(\zeta)\over dx(\zeta)}
\left(
\sum_k t_k \Res_{p\to\infty_-} x^k(p) B(p,\zeta)
- 2i\pi\sum_i \epsilon_i du_i(\zeta) - t dS(\zeta)\right)
\eeq

The equation ${\d \mu\over d a_\alpha} =0$ implies that the differential form
$\sum_k t_k \Res_{p\to\infty_-} x^k(p) B(p,\zeta) - 2i\pi\sum_i \epsilon_i du_i(\zeta) - t dS(\zeta)$ (which clearly has no poles at the branch points), must vanish at all branch points,
and thus is proportional to $dx$.
Let us write it:
\beq
\om(p)dx(p) = \sum_k t_k \Res_{q\to\infty_-} x^k(q) B(p,q) - 2i\pi\sum_i \epsilon_i du_i(p) - t dS(p).
\eeq
Notice that
\beq
B(x,x')=\frac{dx\,\,dx'}{2(x-x')^2} + \frac{1}{\sqrt{\sigma(x)}}\,\times\,{\rm rational\,function\,of\,}x
\eeq
\beq
dS(x)=\frac{1}{\sqrt{\sigma(x)}}\,\times\,{\rm rational\,function\,of\,}x
\eeq
\beq
du_i(x)= \frac{1}{\sqrt{\sigma(x)}}\,\times\,{\rm rational\,function\,of\,}x
\eeq
so that:
\beq
\om(x) = \frac{V'(x)}{2} + \sqrt{\sigma(x)}\,\times\,{\rm rational\,function\,of\,}x.
\eeq
This implies that $\om(x)$ is solution of an algebraic equation of the form
\beq
\om^2(x) - V'(x)\om(x)+P(x)=0
\eeq
where $P(x)$ is some rational function.

Moreover, notice that $\Res_{q\to\infty_-} x^k(q) B(p,q)$ has a pole only when $p\to\infty_-$, i.e. it converges when $p\to\infty_+$ in the first sheet (it diverges in the second sheet), this implies that its contribution to $\om(x)$ is $O(1/x^2)$ as $x\to\infty$.
Similarly, $du_i(x)$ has no pole, so the contribution $du_i/dx$ to $\om$ is $O(1/x^2)$ as $x\to\infty$.
The term $dS(p)$ behaves like $\pm dx/x$ at large $p\to \infty_\pm$.
All this implies that $P(x) $ has no other pole than $x=\infty$, i.e. it is a polynomial, and $\om(x)\sim t/x$ at large $x$.

Moreover we have by definition $\oint_{x'\in\acycle_i} B(x,x')=0$, $\oint_{x\in\acycle_i} dS(x)=0$, $\oint_{x\in\acycle_i} du_j(x)=\delta_{i,j}$, so that
\beq
\oint_{x\in\acycle_i} \om(x)dx=-2i\pi \epsilon_i.
\eeq

Therefore we have proved that the equations $\d\mu/\d a_\alpha=0$ imply that there exists a function $\om(x)$ solution of
\beq
\left\{
\begin{array}{l}
 \om(x)^2-V'(x)\om(x)+P(x)=0 \cr
 \om(x)\sim_{\infty} t/x +O(1/x^2) \cr
 \oint_{\acycle_i} \om dx = - 2i\pi\epsilon_i \cr
\end{array}
\right.
\eeq
i.e. $\om(x)$ is a solution to the loop equation \eq{loopeq1M1}.

\smallskip
{\bf Converse:}

Now assume that $\om$ is solution to loop equations, then it is of the form
\beq
\om =\frac{V'(x)}{2} + \sqrt{\sigma(x)}\,\times\,{\rm polynomial\,of}\,x.
\eeq
One thus sees that
\beq
r(x)=\om(p)dx(p) - \sum_k t_k \Res_{q\to\infty_-} x^k(q) B(p,q) + 2i\pi\sum_i \epsilon_i du_i(p) + t dS(p).
\eeq
is a meromorphic differential form on the Riemann surface of the form $C(x)/\sqrt{\sigma(x)}\,dx$ where $C(x)$ is some polynomial of $x$.
It is easy to see that this polynomial $C(x)$ must behave at most like $O(x^{s-1})$ so that
\beq
C(x) = \sum_i c_{i}\,L_i(x),
\eeq
i.e.
\beq
r(x) = \sum_i c_i du_i(x)
\eeq
and one has $\oint_{\acycle_\alpha} r(x)=0$ so that $c_i=0$, and thus
\beq
r(x)=0.
\eeq

This implies that
\beq
{\d \mu\over d a_\alpha}
= - \Res_{\zeta\to \zeta(a_\alpha)} {dS(\zeta)\over dx(\zeta)}
\left( \om(\zeta) dx(\zeta)\right) 
= - \Res_{\zeta\to \zeta(a_\alpha)} dS(\zeta)\om(\zeta) =0
\eeq
since there is no pole at $\zeta(a_\alpha)$.

This proves the theorem.

}

\subsection{Example: 1-cut case, $s=0$}

The previous variational problem can be further simplified in the genus zero case (1 cut, $s=0$).
For any $\alpha$ and $\gamma$, consider the function $x:\mathbb C^*\to \mathbb C$ defined as:
\beq
x(p)=\alpha+\gamma\left( p+\frac{1}{p}\right)
\eeq
and consider the function:
\beq
\mu(\{t_i\},t;\alpha,\gamma) = \Res_{p\to\infty} V(x(p))\,\frac{dp}{p} - 2t\ln \gamma
\eeq
We have
\beq
\frac{\d\mu}{\d\alpha} = \Res_{p\to\infty} V'(x(p))\,\frac{dp}{p}
\eeq
\beq
\frac{\d\mu}{\d\gamma}  = \Res_{p\to\infty} V'(x(p))\,\left(p+\frac{1}{p}\right)\,\frac{dp}{p} - \frac{2 t}{\gamma}
\eeq
Let us write:
\beq
V'(x(p)) = \sum_{k=0}^{\deg V'} u_k (p^k+p^{-k})
\eeq
The equations $\d\mu/\d\alpha=0$ and $\d \mu/\d\gamma=0$ imply:
\beq
u_0=0
\virg
u_1 = \frac{t}{\gamma}
\eeq

Then, the function:
\beq
\om(p) :=  \sum_{k=1}^p u_k p^{-k}
\eeq
is such that
\beq
V'(x(p))-\om(p) = \sum_{k=1}^p u_k p^{k}
\eeq
and thus
\beq
(V'(x(p))-\om(p))\om(p) 
\eeq
is a polynomial of $p$ and $1/p$ which is symmetric when $p\to 1/p$, i.e. it is a polynomial of $p+1/p$, and so can be written as a polynomial of $x(p)$:
\beq
(V'(x(p))-\om(p))\om(p) = P(x(p)).
\eeq
Moreover the condition $u_1=t/\gamma$ implies that at $p\to \infty$ one has
\beq
\om(p) \sim t/x(p) +O(1/x(p)^2).
\eeq
I.e. we get the loop equations of the 1-matrix model.

\subsection{Link with the free energy}

The free energy is the limit
\beq
F_0 = \lim_{N\to\infty} \frac{t^2}{N^2}\ln Z
\eeq
where $Z$ is the partition function \refeq{defZ1MM}.
It is well known \cite{ZJDFG} that it is worth
\beq
F_0 = \frac{1}{2}\Big( \Res_{p\to\infty_+} V(x(p))\,\om(p) + t\mu^* + \sum_\alpha \epsilon_\alpha\,\oint_{\bcycle_\alpha} \om \Big)
\eeq
where $\om$ is the solution of loop equations, and $\mu^*$ is the value of the functional $\mu$ at its extremum.
It is also well known that:
\beq\label{dF0dtmu}
\frac{\d F_0}{\d t} = \mu^*.
\eeq
so that $\mu^*$ is the value of the derivative of the free energy with respect to $t$.
It can be called the "entropy".

\medskip

When the eigenvalues are real and $V$ is real, i.e. when $\om$ is the Stieljes transform of a positive measure $d\rho$ on $\mathbb R$, extremum of ${\cal S}[d\rho]$ it is known that we have
\beq
F_0 = -\,{\cal S}[d\rho^*].
\eeq

\medskip

\subsubsection{Extremal filling fractions}

Often the filling fractions $\epsilon_\alpha$ are not fixed, and one determines the filling fractions by requiring:
\beq
\frac{\d \Re\,F_0}{\d \epsilon_\alpha}=0
\eeq
i.e.
\beq
\Re\,\oint_{\bcycle_\alpha} \om = 0
\eeq

Then notice that if $\epsilon_\alpha\in \mathbb{R}$ one has
\beq
\Re\,\oint_{\acycle_\alpha} \om  = \Re\, 2\ii\pi\, \epsilon_\alpha = 0
\eeq
and if $t$ is real one has
\beq
\Re\,\oint_{\infty_\pm} \om = \pm \Re\,2\ii\pi\,t =0
\eeq
This implies that for any closed cycle $C$ on the Riemann surface one has
\beq
\Re\,\oint_C \om =0
\eeq
This is the "Boutroux property".

\bd
An algebraic curve has the Boutroux property, iff there exists a one-form $\om$, such that for all closed contour $C$ one has
\beq
\Re\,\oint_C \om =0.
\eeq
In this case, the primitive $h(x) = \Re\,\int_{.}^x \om$, is a harmonic function globally defined on the algebraic curve (indeed the value of $h$ is independent of the choice of integration contour).

\ed

An important property of $F_0$ is that:
\beq
\frac{\d^2 F_0}{\d\epsilon_\alpha \d\epsilon_\beta} = 2\ii\pi\,\oint_{\bcycle_\alpha} du_\beta := 2\ii\pi\,\tau_{\alpha,\beta}
\eeq
and the $s\times s$ matrix $\tau$, called the Riemann matrix of periods, has the well known property \cite{Farkas, Fay} that:
\beq
\tau=\tau^t
\virg
\Im\,\tau>0.
\eeq
Since the imaginary part is positive definite, we have that:
\beq
\Re\,\,
\frac{\d^2 F_0}{\d\epsilon_\alpha \d\epsilon_\beta}
= - 2\pi \Im\,\tau <0
\eeq
i.e. $\Re\,F_0$ is a concave\footnote{Here we have a concave function because we defined $Z=\ee{F}$ instead of the usual Gibbs convention $Z=\ee{-{\cal F}}$ with which ${\cal F}=-F$ is convex.} function of filling fractions, and thus it has a unique maximum.

So, in case the filing fractions were not fixed at the beginning, they are chosen as
the ones which maximize $\Re\, F_0$.

\section{The 2 matrix model}

A similar variational principle can be found for the loop equations of the 2-matrix model \cite{Kazakov}.

\subsection{Introduction 2-matrix model}

Consider two random hermitian matrices (or two random normal matrices with eigenvalues on some contours) $M_1, M_2$ of size $N$, with probability law:
\beq
\frac{1}{Z}\,\ee{-\frac{N}{t}\,\Tr \left(V_1(M_1)+V_2(M-2)-M_1 M_2\right)}\,\,dM_1\,dM_2
\eeq
where $V_1(x) = \sum_k \frac{t_k}{k} x^k$, and$V_2(y) = \sum_k \frac{\td t_k}{k} y^k$ are polynomials called the potentials, and $t>0$ is often called "temperature", and $Z$ is the partition function:
\beq
Z=\int_{H_N\times H_N}\,\ee{-\frac{N}{t}\,\Tr \left(V_1(M_1)+V_2(M-2)-M_1 M_2\right)}\,\,dM_1\,dM_2 .
\eeq

The expectation value of the resolvent of matrix $M_1$:
\beq
W(x)  = \frac{t}{N}\,\mathbb E \left( \Tr \,(x-M_1)^{-1} \right)
\eeq
plays an important role, indeed it encodes the information on the spectrum of $M_1$.

In many cases (depending on the choice of potentials $V_1, V_2$, and on the choices of contours), it is known or conjectured (see \cite{ZJDFG} for instance), that $W(x)$ has a large $N$ limit, which we write:
\beq
\exists\,\, \mathop{{\lim}}_{N\to\infty} \,\, V'_1(x)-W(x)  =  \om(x)
\eeq
and in many cases (again depending on the choice of potentials $V_1, V_2$ and contours), it is an algebraic function of $x$, i.e. it satisfies an algebraic equation \cite{Kazakov, KazMar, staudacher, eynm2m}:
\beq
P(x,\om(x))=0
\qquad ,\,\,
P(x,y) = \sum_{i,j} P_{i,j} x^i\,y^j.
\eeq

For the 2-matrix model, the polynomial $P(x,y)$ is in general not quadratic in $y$, instead it takes the form \cite{eynm2m}:
\beq
P(x,y) = (y-V'_1(x))\,(x-V'_2(y)) + Q(x,y)
\eeq
where $Q(x,y)$ is a polynomial such that:
\beq
\deg_x Q < \deg V'_1
\qquad , \quad
\deg_y Q < \deg V'_2.
\eeq

\subsubsection{Some algebraic geometry}

The equation $P(x,y)=0$ is an algebraic equation, it defines a compact Riemann surface $\curve$.
This Riemann surface has a certain genus $\genus$.

\subsubsection{Filling fractions}

Let us define for $\alpha=1,\dots,\genus$, a basis of $2\genus$ non--contractible cycles on $\curve$:
\beq
\acycle_{\alpha=1,\dots,\genus}
\virg
\bcycle_{\alpha=1,\dots,\genus},
\eeq
with canonical symplectic intersections
\beq
\acycle_\alpha \cap \bcycle_\beta = \delta_{\alpha,\beta}
\virg
\acycle_\alpha \cap \acycle_\beta = \emptyset
\virg
\bcycle_\alpha \cap \bcycle_\beta = \emptyset.
\eeq
Such a canonical basis always exists but is not unique.

Very often, it is interesting to consider matrix models with "fixed filling fractions", i.e. where the number of eigenvalues of $M_1$ or $M_2$ in a certain region of the complex plane is held fixed.
The number $n_\alpha$ of eigenvalues of $M_1$ enclosed by a clockwise contour $C_\alpha$ is:
\beq
n_\alpha = -\,\frac{N}{2i\pi\,t}\,\oint_{C_\alpha}\,W(x)\,dx
\eeq
In the large $N$ limit, the fixed filling fraction condition amounts to fix:
\beq
\frac{t\,n_\alpha}{N} =\,\frac{1}{2i\pi}\,\oint_{\acycle_\alpha}\,\om(x)\,dx = \epsilon_\alpha
\eeq
The numbers $\epsilon_\alpha$ are called "filling fractions", they tell the number (times $t/N$)  of eigenvalues of $M_1$ which concentrate in regions enclosed by the $\acycle_\alpha$'s.

\subsubsection{Loop equations}

Our goal now is to find the polynomial $Q(x,y)$.

It is well known \cite{eynm2m, KazMar} that this polynomial can be  determined by the following equations:

\bd[Loop equations]

The loop equations of the 2-matrix model with potentials $V_1, V_2$ and with filling fractions $\epsilon_\alpha$ is the following set of equations \cite{eynm2m, KazMar, eynm2mg1}:
\beq\label{loopeq2M}
\encadremath{
\left\{\begin{array}{l}
\exists\,{\rm polynomial}\, Q(x,y)\,\,\, {\rm such\,that}\,\, (\om(x)-V'_1(x))\,(x-V'_2(\om(x))) + Q(x,\om(x)) = 0 \cr
{} \cr
\om(x)\mathop{{\sim}}_{\infty_+} V'_1(x)- t/x +O(1/x^2) \cr
{} \cr
x\mathop{{\sim}}_{\infty_-} V'_2(\om(x)) - t/\om(x) +O(1/\om(x)^2) \cr
{} \cr
\forall\,\alpha=1,\dots,\genus\, , \qquad  -\,\frac{1}{2i\pi}\,\oint_{\acycle_\alpha}\,\om(x)\,dx = \epsilon_\alpha
\end{array}\right.
}\eeq

\ed

Let us check that this system implies as many equations as unknowns.
The 2 equations regarding the behaviors at $\infty_\pm$ imply that $\deg_x Q < \deg V'_1$ and $\deg_y Q < \deg V'_2$, and they also imply that the leading term (largest power of both $x$ and $y$) is of the form:
\beq
Q(x,y) \sim t\,\frac{V'_1(x)\,V'_2(y)}{xy}.
\eeq
This implies that the number of unknown coefficients of $Q(x,y)$ is $\deg V'_1 \times \deg V'_2 -1$, which is also\footnote{classical result of algebraic geometry, the genus is the number of interior points of the Newton's polygon. And here the Newton's polygon has $\deg V'_1 \times \deg V'_2 -1$ interior points.} the genus $\genus$ of the Riemann surface of equation $P(x,y)=0$.
Therefore the number of unknown coefficients of $Q(x,y)$ matches the number of filling fraction conditions.

\medskip
Our goal is not to study those equations, in particular their number of solutions (existence or unicity questions), as there already is a large literature about them, but to show that the same set of equations \eq{loopeq2M} can be obtained from a variational principle.

\subsection{Algebro-geometric notations}

Let $\curve$ be a compact Riemann surface of genus $\genus$, defined by an algebraic equation $P(x,y)=0$.

This means that every point $p\in \curve$ corresponds to a point $(x(p),y(p))\in \mathbb C^2$ such that $P(x(p),y(p))=0$.
In other words there exists two analytical meromorphic functions $x:\curve \to \mathbb C$, $y:\curve\to\mathbb C$
\beq
\left\{\begin{array}{ll}
x: \quad &  \curve \to \mathbb C \cr
& p \mapsto x(p)
\end{array}\right.
\virg
\left\{\begin{array}{ll}
y: \quad &  \curve \to \mathbb C \cr
& p \mapsto y(p)
\end{array}\right.
\eeq
such that
\beq
\{(x,y)\in\mathbb C^2\,|\,\,P(x,y)=0\} \quad \equiv \quad \{(x(p),y(p))\,|\,p\in \curve\}.
\eeq

\subsubsection{Branchpoints}

We define branchpoints as the zeroes of the differential $dx$ on $\curve$:
\beq
dx(e_\alpha)=0.
\eeq
Their $x$--projection is denoted:
\beq
a_\alpha=x(e_\alpha).
\eeq
We assume that, generically, those zeroes are simple zeroes, i.e. a good local coordinate on $\curve$ near $e_\alpha$ is:
\beq
\zeta = \sqrt{x-a_\alpha} \virg x=a_\alpha+\zeta^2 \quad ,\, dx=2\zeta d\zeta.
\eeq

\subsubsection{Holomorphic forms}

There exists \cite{Fay,Farkas}) a unique basis of holomorphic forms $du_i(p)$ on $\curve$ normalized on $\acycle$-cycles such that:
\beq
\oint_{\acycle_i} du_j(p) = \delta_{i,j}
\virg i,j=1,\dots,\genus.
\eeq
One can always write:
\beq
du_i(p) = \frac{R_i(x(p),y(p))\,\,dx(p)}{P'_y(x(p),y(p))}
\eeq
where $R_i(x,y)\in \mathbb C[x,y]$ is the unique polynomial of degree $\deg_x R_i<\deg V'_1$ and $\deg_y R_i<\deg V'_2$, chosen such that $du_i(p)$ has no pole on $\curve$ and $\oint_{\acycle_i} du_j(p) = \delta_{i,j}$.

\subsubsection{2nd kind form}

Similarly, there exists a unique symmetric bi--differential form $B(x,y)\in T^*(\curve)\otimes T^*(\curve)$, having a double pole on the diagonal, and no other pole, and normalized on $\acycle$-cycles:
\beq
B(p,p') \mathop{\sim}_{p\to p'} \frac{d\zeta(p)\otimes d\zeta(p')}{(\zeta(p)-\zeta(p'))^2} + {\rm analytical\,at\,}p=p'
\eeq
\beq
\forall\,i=1,\dots,\genus\, ,\,\,\forall\,p\in\curve\,\,\, \qquad \oint_{p'\in \acycle_i} B(p,p')=0
\eeq
$B(p,p')$ is called the "fundamental form of the second kind"  or (derivative of) "Green--function" or "heat kernel" on $\curve$.

It has the property \cite{Fay} that:
\beq
\oint_{p'\in \bcycle_i} B(p,p') = 2\ii\pi du_i(p).
\eeq

We also define the 3-rd kind differential:
\beq
dS(p) = \int_{p'=\infty_-}^{\infty_+} B(p,p')
\eeq
where the integration path is chosen\footnote{Notice that $\curve\setminus \cup_\alpha \acycle_\alpha \cup_\alpha \bcycle_\alpha$ is simply connected, and thus $dS$ is well defined.} such that it doesn't intersect any $\acycle$-cycle or $\bcycle$-cycle.

Then, let $p_0$ be an arbitrary basepoint and define
\beq
\L(p)=\exp{\int_{p_0}^p dS}
\eeq
where again the integration contour avoids $\acycle$-cycles and $\bcycle$-cycles.
Let
\beq
\gamma = \lim_{p\to\infty_+} x(p)^{1/\deg_{\infty_+}(x)}\,/\L(p)
\eeq
\beq
\td\gamma =  \lim_{p\to\infty_-} \L(p) /\, y(p)^{1/\deg_{\infty_-}(y)}
\eeq
Notice that the product $\gamma{\td\gamma}$ is independent of the choice of $p_0$.

%
%
%

\subsection{The variational principle}

\bd

Consider the following functional:
\bea
\mu(\{t_k\},\{\td t_k\},t;(\curve,x,y))
&:=& \sum_k t_k \Res_{p\to\infty_+} x(p)^k dS(p)
- \sum_k \td{t}_k  \Res_{p\to\infty_-}  y^k dS(p) \cr
&& - c \Res_{p\to \infty_+} x(p)y(p)\, dS(p)
+ \sum_i \epsilon_i \oint_{\bcycle_i} dS(p)
- t\,\ln{\gamma \td\gamma} \cr
\eea
where $(\curve,x,y)$ is a compact Riemann surface of genus $\genus$ with 2 distinct marked points called $\infty_+$ and $\infty_-$,  and $x$ and $y$ any two meromorphic functions on $\curve\to \mathbb P^1$.

\ed

It is such that an extremum of $\mu$, i.e. $d\mu=0$ is a solution of the loop equation \eq{loopeq2M}.

\bt
The set of equations (differential with respect to variations of $({\cal C},x,y)$)
\beq
d\mu=0
\eeq
is equivalent to the loop equations \eq{loopeq2M}.

\et

\proof{
Let $(\curve,x,y)$ be a compact Riemann surface of genus $\genus$, with 2 marked points $\infty_\pm$, and $x$ and $y$ any two meromorphic functions on $\curve\to \mathbb P^1$.

The tangent (infinitesimal variations) of the moduli space of $(\curve,x,y)$ is isomorphic to the space of meromorphic forms on $\curve$.
Notice that one can vary at the same time the complex structure of $\curve$, as well as the functions $x$ and $y$.

Let $\delta$ denote  a tangent direction, i.e.
\beq
\delta(y)dx - \delta(x)dy = \Omega
\eeq
$\Omega$ is a meromorphic form.

The Rauch variational formula gives:
\beq
\delta\,\left. B(p,q)\right|_{x(p),x(q)} = \sum_\alpha \Res_{s\to e_\alpha} {B(p,s)B(q,s)\Om(s)\over dx(s)dy(s)}
\eeq
thus:
\beq
\delta\,\left. dS(p)\right|_{x(p)} = \sum_\alpha\Res_{s\to e_\alpha} {B(p,s)dS(s)\Om(s)\over dx(s)dy(s)}
\eeq
\beq
\delta\,\left. \ln\L(p)\right|_{x(p)} =\sum_\alpha \Res_{s\to e_\alpha} {dE_p(s)dS(s)\Om(s)\over dx(s)dy(s)}
\eeq
\beq
\delta\, \ln\gamma  = - \Res_{s\to e_\alpha} {dS_{\infty_x,o}(s)dS(s)\Om(s)\over dx(s)dy(s)}
\eeq
By the chain rule we have:
\bea
\delta\,\left. dS(p)\right|_{y(p)}
&=& \delta\,\left. dS(p)\right|_{x(p)} -d\left({\Om(p)dS(p)\over dx(p)dy(p)}\right)
\eea
\beq
\delta\,\left. dS(p)\right|_{y(p)} = \Res_{s\to e_\alpha} {B(p,s)dS(s)\Om(s)\over dx(s)dy(s)}
\eeq
\beq
\delta\,\left. \ln\L(p)\right|_{y(p)} = \Res_{s\to e_\alpha} {dE_p(s)dS(s)\Om(s)\over dx(s)dy(s)}
\eeq
\beq
\delta\, \ln\td\gamma  = \Res_{s\to e_\alpha} {dS_{\infty_y,o}(s)dS(s)\Om(s)\over dx(s)dy(s)}
\eeq
and
\beq
\delta\, \ln(\gamma\td\gamma)  = - \Res_{s\to e_\alpha} {dS(s)dS(s)\Om(s)\over dx(s)dy(s)}
\eeq

Thus:
\bea
\delta\mu
&=& \Res_{e} {\Om dS\over dx dy}(-cydx+ \sum t_k \Res_{\infty_+} x^k B -  \sum \td{t}_k \Res_{\infty_y} y^k B + c \Res_{\infty_y} xy B \cr
&& +\sum \epsilon_i du_i + t dS) \cr
\eea

$\delta\mu=0$ for any meromorphic 1-form $\Omega$ implies that 
\beq
cydx = \sum t_k \Res_{\infty_x} x^k B -  \sum \td{t}_k \Res_{\infty_y} y^k B + c \Res_{\infty_y} xy B +\sum \epsilon_i du_i + t dS
\eeq
This expression of $ydx$ implies that near $\infty_+$ one has
\beq
cy\sim V'_1(x) - \frac{t}{x} + O(1/x^2)
\eeq
Doing the same computation with fixed $y$ instead of fixed $x$ yields:
\beq
cxdy = \sum \td t_k \Res_{\infty_-} y^k B -  \sum t_k \Res_{\infty_+} x^k B + c \Res_{\infty_+} xy B -\sum \epsilon_i du_i - t dS
\eeq
which gives that near $\infty_-$ one has
\beq
cx\sim V'_2(y) - \frac{t}{y} + O(1/y^2)
\eeq
and moreover
\beq
\oint_{\acycle_i} ydx = 2i\pi\,\epsilon_i.
\eeq
The reverse proposition is obvious, this concludes the proof.

}

\subsection{Example: Genus zero curves}

Genus 0 curves can be parametrized by rational functions.
Consider $(\curve,x,y)$ where $\curve$ is a genus zero curve with 2 marked points, i.e. it is the Riemann sphere $\mathbb P^1$, and we can chose the 2 marked points to be $\infty_+=\infty$ and $\infty_-=0$, and $x$ and $y$ are 2 rational functions.
Let us assume that $x$ has a simple pole at $p=\infty$ and an arbitrary pole at $p=0$, and $y$ has a simple pole at $p=0$ and an arbitrary pole at $p=\infty$:
\bea
x(p) &=& \sum_{k=-1}^{d_2} \alpha_k p^{-k} \cr
y(p) &=& \sum_{k=-1}^{d_1} \beta_k p^{k}
\eea

Consider the following function:
\beq
\mu(\{t_i\},\{\td{t}_i\},c,t;\{\alpha_k\},\{\beta_k\})
\eeq
\bea
\mu
&:=& \sum_k t_k \Res_{\infty} x(p)^k {dp\over p}
+ \sum_k \td{t}_k \Res_{\infty} y(p)^k {dp\over p} \cr
&& - c \Res_{\infty} x(p)y(p)\, {dp\over p}
- t\ln{(\alpha_{-1}\beta_{-1})} \cr
\eea

We have:
\bea
{\d\mu\over \d \alpha_j}
&=&  \Res_{\infty} (\sum_k k t_k x(p)^{k-1}-cy(p)) {p^{-j} dp\over p} - t{\delta_{j,-1}\over \alpha_{-1}}
\eea
${\d\mu\over \d \alpha_j}=0$  implies:
\beq
\forall\,j=-1,\dots,d_2 \quad {\d\mu\over \d \alpha_j}=0 \quad \longrightarrow\quad cy(p) = \sum_k k t_k x(p)^{k-1} - {t\over x(p)} +O(1/p^2)
\eeq
and similarly with the $\beta_j$'s
\beq
\forall\,j=-1,\dots,d_1 \quad{\d\mu\over \d \beta_j}=0 \quad \longrightarrow\quad cx(p) = \sum_k k \td{t}_k y(p)^{k-1} - {t\over y(p)} +O(p^2)
\eeq
i.e. we obtain the loop equations, for instance as written in \cite{eynchain}.

\section{Generalization: algebraic plane curve with fixed behaviors at poles\label{secalgcurve}}

The 1-matrix and 2-matrix loop equations are special cases of the following problem (related to the Witham hierarchy \cite{Kri, MarcoF}):

\medskip

{\bf Problem:}
Let $\genus$, $m$, $\{t_{k,j}\}_{k=1,\dots,m, \, j=1,\dots,d_k}$, $\{\epsilon_i\}_{i=1,\dots,\genus}$, $\{X_j\}_{j=1,\dots,m}$ be given.

\smallskip

Find $(\curve,x,y)$ where $\curve$ is a compact Riemann surface of genus $\genus$, with $m$ marked points $\{\infty_k\}_{k=1,\dots,m}$, and with $2\genus$ closed cycles whose homology class form a  symplectic basis of cycles $\acycle_i\cap \bcycle_j=\delta_{i,j}$, and $x$ and $y$ are 2 meromorphic functions on $\curve$, such that:

$\bullet$ $y$ and $x$ are holomorphic on $\curve\setminus\{\infty_k\}_{k=1,\dots,m}$,

$\bullet$ 
\beq
\forall\,k=1,\dots,m \qquad  x(\infty_k)=X_k.
\eeq
If $X_k=\infty$ we define the local coordinate $\zeta_k(p) = x(p)^{-1/\deg_{\infty_k}(x)}$, and if $X_k\neq \infty$ we define $\zeta_k(p) = x(p)-X_k$.

$\bullet$ the 1-form $ydx$ has a prescribed negative part of its Laurent series expansion near $\infty_k$:
\beq
y(p)dx(p)\mathop{{\sim}}_{\infty_k} \sum_{j= 0}^{d_k} t_{k,j} \,\zeta_k(p)^{-j-1}\,d\zeta_k(p)  + {\rm analytical\,at\,}\infty_k
\eeq

$\bullet$ one has prescribed filling fractions
\beq
\frac{1}{2i\pi}\oint_{\acycle_i} ydx= \epsilon_i.
\eeq

\medskip

Here we shall not consider the question of existence and/or unicity of a solution. We just mention that a necessary condition for a solution to exist is that the sum of residues of a meromorphic form vanishes i.e.
\beq
\sum_k t_{k,0}=0.
\eeq

From now on, we assume that this condition is fulfilled, and we shall merely reformulate the question as a variational principle.

\subsection{Variational principle}

\bd
Let $(\curve,x)$ be a Hurwitz space, where $\curve$ is a Riemann surface of genus $\genus$, with marked points $\infty_k$, and with a given symplectic basis of cycles $\acycle_i\cap \bcycle_j=\delta_{i,j}$, and $x$  is a meromorphic function on $\curve$, used as a projection on the base Riemann sphere: $x: \curve\to \overline{\mathbb C}$.

We define $\forall\,i,i'$ any two distinct $\infty_i\neq \infty_{i'}$:
\bea
\mu_{i,i'}(\{t_{k,j}\};(\curve,x,y))
&=& \sum_k \Res_{p\to\infty_k} \sum_{j=1}^{d_k} \frac{t_{k,j}}{j}\,\zeta_j(p)^{-j}\,dS_{\infty_i,\infty_{i'}}(p) \cr
&& + \sum_k t_{k,0} \ln\gamma_k
+ \sum_\alpha \epsilon_\alpha \oint_{p\in\bcycle_\alpha} dS_{\infty_i,\infty_{i'}}(p) \cr
\eea
where
\beq
dS_{\infty_{i},\infty_{i'}}(p) = \int_{\infty_{i}}^{\infty_{i'}} B(.,p)
\eeq
and if $o$ is an arbitrary generic point of $\curve$
\beq
\gamma_k = \frac{E(\infty_i,\infty_k)\,E(\infty_{i'},o)}{E(\infty_{i'},\infty_k)\,E(\infty_i,o)}.
\eeq
notice that since $\sum_k t_{k,0}=0$, we have that $\sum_k t_{k,0}\ln\gamma_k$ is independent of the choice of $o\in\curve$.
\ed

\bt
For any $i,i'$, let $\mu=\mu_{i,i'}$, then a solution of $d\mu=0$ is a solution to the problem above.
\et

\proof{
The tangent space to the moduli space of $(\curve,x,y)$, is the space of meromorphic forms $\Omega$ on $\curve$ such that:
\beq
\delta y\,dx - x\delta y = \Omega
\eeq
Moreover, if we consider that $x$ and $y$ have poles only at the $\infty_k$'s, we require that $\Omega$ can have poles only at the $\infty_k$'s.

As before, we use Rauch formula and get:
\bea
\frac{\d \mu}{\d a_\alpha}
&=& 
\Res_{p\to e_\alpha} \frac{dS_{\infty_i,\infty_{i'}}(p)}{dx(p)}\,\Big(
\sum_k \sum_{j\geq 1} \frac{t_{k,j}}{j} \Res_{q\to \infty_k} B(p,q) \zeta_k(q)^{-j} \cr
&&  + \sum_k t_{k,0} dS_{\infty_k,o}(p) + 2\ii \pi\sum_\alpha \epsilon_\alpha du_\alpha(p) 
\Big) \cr
\eea
(notice again that since $\sum_k t_{k,0}=0$, then $\sum_k t_{k,0} dS_{\infty_k,o}$ is independent of the choice of $o\in\curve$).

Notice that the quantity inside the bracket has no pole at $e_\alpha$, and thus the fact that the residue vanishes implies that the quantity in the bracket vanishes at $e_\alpha$, and thus can be divided by $dx$:
\bea
y
&=&\frac{1}{dx}\,\Big(
\sum_k \sum_{j\geq 1} \frac{t_{k,j}}{j} \Res_{q\to \infty_k} B(p,q) \zeta_k(q)^{-j} \cr
&&  + \sum_k t_{k,0} dS_{\infty_k,o}(p) + 2\ii \pi\sum_\alpha \epsilon_\alpha du_\alpha(p) 
\Big)
\eea
is a meromorphic function with the required Laurent series behavior near poles and filling fractions, it is thus a solution to the problem.

}

\section{Conclusion}

We have seen that the loop equations of various matrix models, which consist in finding a plane curve with prescribed asymptotic behaviors at poles and prescribed filling fractions on $\acycle$-cycles, are equivalent to a local variational principle.

Contrarily to the energy functional ${\cal S}$ or $F_0$, the functional $\mu$ doesn't have convexity properties, so one cannot easily conclude to the existence of a solution of the variational principle. However, the functional $\mu$ is in fact easier to compute, and the loop equations easier to derive from $\mu$.
Also, the geometric meaning of that $\mu$ needs to be understood, in particular the equation \eq{dF0dtmu}.

\medskip

In this article we have explicitly considered only the 1 and 2-matrix models, although section \ref{secalgcurve} guarantees that it also applies to the "chain of matrices" \cite{eynchain, eynmultimat} matrix model, and possibly more. Also, we have written the explicit proof for 1 and 2 matrix model only for polynomial potentials, and again section \ref{secalgcurve} guarantees that the same works for potentials whose derivative is a rational function (called semi-classical potentials \cite{MarcoF}), or also for matrix models with hard edges \cite{MarcoF, eylooprat}.

\section*{Acknowledgments}
This work  is partly supported by the 
Quebec government with the FQRNT, and the Field-Knot ERC grant of P. Sulkowski.



\end{document}